\documentclass[prb,twocolumn,showpacs,showkeys,superscriptaddress,amsmath,amssymb]{revtex4}
\usepackage{graphicx}
\usepackage{multirow,slashbox}
\DeclareTextSymbol{\degre}{T1}{6}
\DeclareTextSymbol{\degre}{OT1}{23}

\begin{document}

\title{Magnetic structure and charge ordering in Fe$_3$BO$_5$ ludwigite}

\author{P. Bordet}
\email{pierre.bordet@grenoble.cnrs.fr}
\affiliation{Institut NEEL\\
CNRS \& UJF, BP 166, 38042 Grenoble Cedex 9, France}

\author{E. Suard}
\affiliation{Institut Laue Langevin\\
 BP 156, 38042 Grenoble Cedex 9, France}

%\date{today}

\begin{abstract}
The crystal and magnetic structures of the three-leg ladder compound Fe$_3$BO$_5$ have been investigated by  single crystal x-ray diffraction and neutron powder diffraction. $Fe_3BO_5$ contains two types of three-leg spin ladders. It shows a charge ordering transition at 283 K, an antiferromagnetic transition at 112 K, ferromagnetism below 70 K and a weak ferromagnetic behavior below 40K. The x-ray data reveal a smooth charge ordering and an incomplete charge localization down to 110K. Below the first magnetic transition, the first type of ladders orders as ferromagnetically coupled antiferromagnetic chains, while below 70K the second type of ladders orders as antiferromagnetically coupled ferromagnetic chains. 
\end{abstract}


\maketitle

\section{Introduction\label{sec:intro}}
\begin{figure}[btp]
	\begin{center}
		\includegraphics[width=0.25\textwidth]{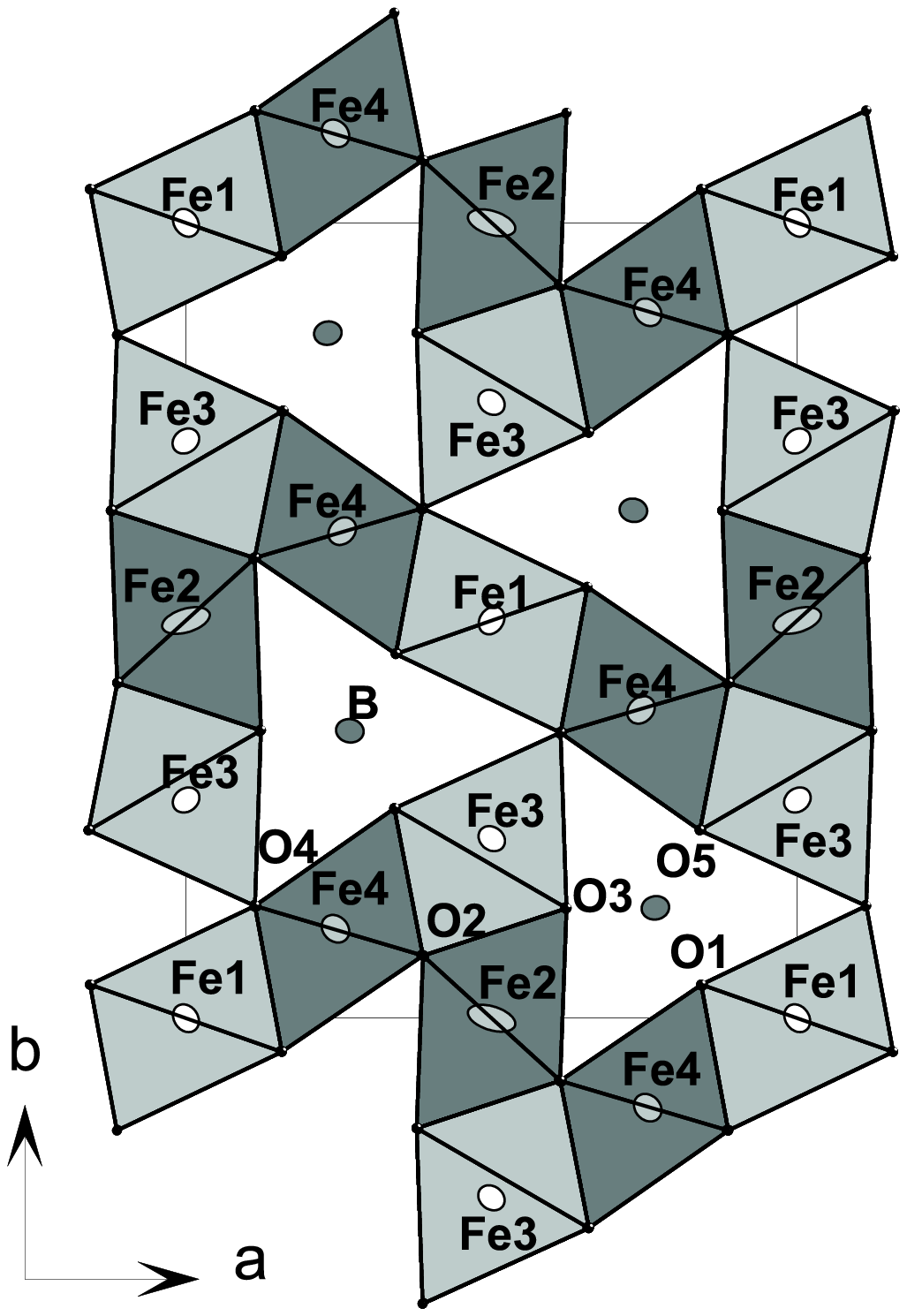}
		\caption{Structure of the high temperature phase of Fe$_3$BO$_5$ above 290K projected along the c axis. Light gray octahedra contain Fe cations with a +2 valence, dark gray ones Fe cations with an average valence of +2.66.}
		\label{HTlabels}
	\end{center}
\end{figure}

The so-called homometallic ludwigite Fe$_3$BO$_5$ is a mixed valence low dimensional semiconductor which shows a number of exciting electronic and magnetic properties \cite{conti,guimar}. The general chemical formula for ludwigites is $M_2^{2+}M^{3+}O_2BO_3$ where $M^{2+}$ or $M^{3+}$ are 3d transition metal ions. The $M^{2+}$ and $M^{3+}$ ions are located at the centers of edge sharing oxygen octahedra forming zig-zag walls. The crystal structure of the Fe$_3$BO$_5$ high temperature phase is depicted in Figure \ref{HTlabels}, where the atom labeling is shown. The symmetry is orthorhombic, space group  $Pbam$, with a=9.462\AA, b=12.308\AA, c=3.075\AA. The $Fe^{2+}$ and $Fe^{3+}$ cations occupy four distinct metal sites. The quasi-two dimensional structure can be viewed as resulting from the presence of two types of three-leg spin ladders sub-units of $Fe$ cations. Ladders I are built upon the $Fe2$ and $Fe4$ $Fe^{3+}$ cations having localized high-spins S=5/2 with one additional itinerant electron per rung. Ladders II involve the $Fe1$ and $Fe3$ $Fe^{2+}$ cations with S=2 localized spins according to M$\ddot{o}$ssbauer spectroscopy \cite{guimar, larrea}. Ladders I show a structural and charge ordering transition at Tc = 283K such that long and short bonds alternate along the ladder c-axis \cite{mir1}. This transition is accompanied by an anomaly in the magnetization and a change of slope of the resistivity. The origin of this transition has been discussed in terms of excitonic instability \cite{latge}. $Fe_3BO_5$ shows an antiferromagnetic transition at 112K, weak ferromagnetism below 70K and another magnetic transition at 40K where ferromagnetism disappears \cite{conti,guimar}.
It is of primary importance to determine the evolution of charge and magnetic ordering in the whole temperature range between 300K and 10K in order to gain insight into the complex physical behavior of this material. Previous x-ray diffraction studies \cite{mir1,mir2} only provide structural information  at 300K, 144K and 15K. Magnetic structure investigation at 5K was reported by Attfield et al. \cite{attfield}, but nothing is known about magnetic orderings at intermediate temperatures. Here we present the results of a x-ray single crystal diffraction experiment and of a neutron powder diffraction experiment both as function of temperature, aiming at providing detailed information about the charge and spin behavior of this complex and original compound. Preliminary results were given in refs. \cite{bordet1,bordet2}.  

%%%%%%%%%%%%%%%%%%%%%%%%
%====================================
\section{Experimental\label{sec:experimental}}

Single crystal and powder sample used for this work were synthesized and characterized according to ref.\cite{guimar}. 

\subsection{Single crystal x-ray diffraction\label{XRPD}}

\begin{figure}[btp]
	\begin{center}
		\includegraphics[width=0.4\textwidth]{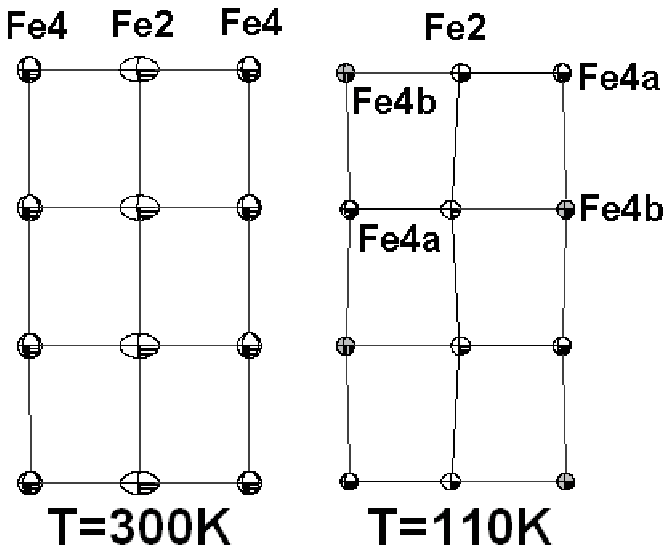}
		\caption{Structural arrangement of the Fe4-Fe2-Fe4 ladders above and below the structural phase transition in Fe$_3$BO$_5$. The vertical direction corresponds to the c axis.}
		\label{ladderhtbt}
	\end{center}
\end{figure}

\begin{figure}[btp]
	\begin{center}
		\includegraphics[width=0.4\textwidth]{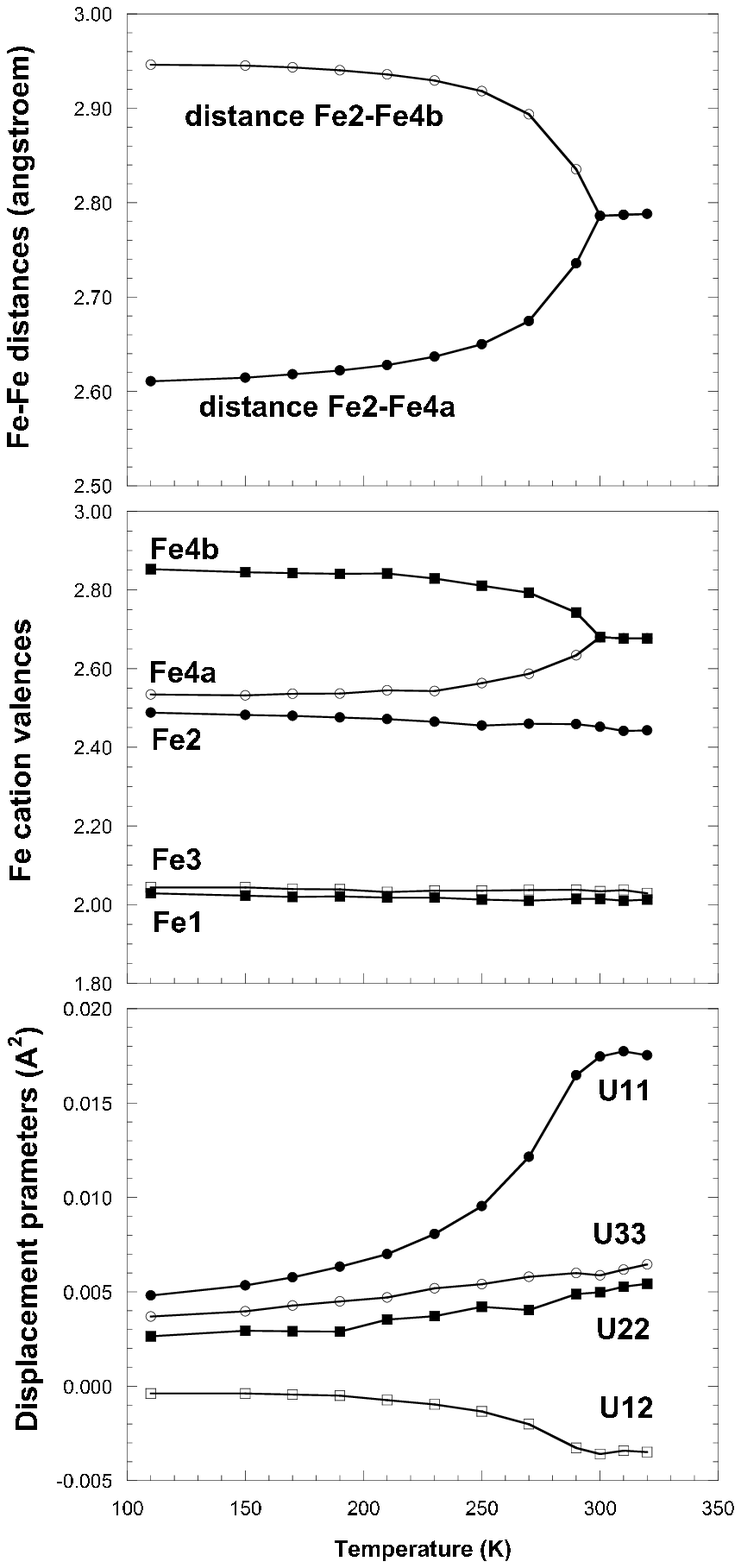}
		\caption{Charge ordering and Fe cation pair formation in Fe$_3$BO$_5$ at the structural phase transition. Top : Fe-Fe distances along rungs of the Fe4-Fe2-Fe4 ladders. Middle : Fe cation valences calculated with the b.v.s. method. Bottom : Atomic displacement parameters for the Fe2 cation. Error bars are smaller than the marker sizes}
		\label{transit}
	\end{center}
\end{figure}

In order to investigate in close detail the charge ordering occurring at 283K, we performed a single crystal x-ray diffraction experiment as a function of temperature to follow the evolution of the structural parameters across the transition and in a large temperature range. The experiment was carried out with a Nonius KappaCCD diffractometer equipped with graphite monochromatized AgK$\alpha$ radiation ($\lambda$=0.5608\AA). The temperature was controlled with an Oxford Cryosystem nitrogen gas blower. A single crystal sample was ground to a sphere of 0.1 mm radius. 180\degre omega scans with 2\degre frame size were used to collect the data up to sin$\theta$/$\lambda$=0.9, with a 35 mm sample to detector distance, every 20K between 320K and 110K on cooling. CCD data were processed, averaged and corrected for absorption with the Denzo-SMN\cite{denzo} and Maxus\cite{maxus} packages, refinements and structure analyzes were performed using the Jana2000 software\cite{jana}. Although the transition was reported to appear at 283K\cite{mir}, we could detect superstructure spots characteristic of a doubling of the c axis parameter already on the 290K data. These data and those at lower temperature were therefore collected for the refinement on the basis of this double cell, while the 300K, 310K and 320K data were refined with the high temperature phase unit cell. For the low temperature phase, each experiment leads to about 2450 unique reflections of which approximately one half were superstructure reflections (l odd), and $\approx$ 1750 were considered as observed, i.e. had $I > 3 \sigma(I)$ (where $I$ and $\sigma(I)$ denote a Bragg reflection intensity and its standard deviation, respectively). For the high temperature phase, about 1260 unique reflections were obtained, of which about 1050 were observed. 
The structure of ludwigite has space group $Pbam$ with cell parameters a=9.465\AA, b=12.310\AA, c=3.077\AA  at 320K. As stated above, all superstructure spots appearing below 290K could be indexed by doubling the c axis parameter. The systematic extinction condition lead to space group $Pbnm$, as reported by Mir et al.\cite{mir}. The refined atomic coordinates at 320K and 110K are shown in Table \ref{xray-data-tab}. 
%%%%%%%%%%%%%%%%%%%%%%%%
%====================================
\begin{table}[!hb]
	\begin{center}
		\begin{tabular}{|c|c|c|c|c|}
		\hline
		\multicolumn{5}{|c|}{T=320K}\\
		\hline
		Atom & Pos. & x & y & z \\
		\hline
		Fe1 & 2a & 0 & 0 & 0 \\ 
		\hline
		Fe2 & 2d & 1/2 & 0 & 1/2 \\
		\hline
		Fe3 & 4g & 0.00029(3) & 0.27433(3) & 0 \\
		\hline
		Fe4 & 4h & 0.74436(3) & 0.38746(3) & 1/2 \\
		\hline
		B & 4h & 0.2687(3) & 0.3617(2) & 1/2 \\
		\hline
		O1 & 4h & 0.8431(2) & 0.0427(1) & 1/2 \\
		\hline
		O2 & 4g & 0.3874(2) & 0.0787(1) & 0 \\
		\hline
		O3 & 4h & 0.6229(2) & 0.1382(1) & 1/2 \\
		\hline
		O4 & 4g & 0.1130(2) & 0.1408(1) & 0 \\
		\hline
		O5 & 4h & 0.8409(2) & 0.2360(1) & 1/2 \\
		\hline
		\multicolumn{5}{|c|}{T=110K}\\
		\hline
		Atom & Pos. & x & y & z \\
		\hline
		Fe1 & 4a & 0 & 0 & 0\\
		\hline
		Fe2 & 4c & 0.51174(3) & -0.00266(2) & 1/4\\
		\hline
		Fe3 & 8d & 0.50017(2) & 0.22596(2) & 0.00257(3)\\
		\hline
		Fe4a & 4c & 0.74924(3) & 0.39071(3) & 1/4\\
		\hline
		Fe4b & 4c & 0.25988(3) & 0.61513(3) & 1/4\\
		\hline
		Ba & 4c & 0.2652(3) & 0.3629(2) & 1/4\\
		\hline
		Bb & 4c & 0.7292(3) & 0.6400(2) & 1/4\\
		\hline
		O1a & 4c & 0.8451(2) & 0.0437(1) & 1/4\\
		\hline
		O1b & 4c & 0.1585(2) & 0.9583(1) & 1/4\\
		\hline
		O2 & 8d & 0.8880(1) & 0.4218(1) & -0.0096(2)\\
		\hline
		O3a & 4c & 0.6255(2) & 0.1387(1) & 1/4\\
		\hline
		O3b & 4c & 0.3809(2) & 0.8628(1) & 1/4\\
		\hline
		O4 & 8d & 0.6133(2) & 0.3591(1) & -0.0025(2)\\
		\hline
		O5a & 4c & 0.8435(2) & 0.2376(1) & 1/4\\
		\hline
		O5b & 4c & 0.1620(2) & 0.7650(1) & 1/4\\
		\hline
		\end{tabular}
	\end{center}
	\caption{Positional parameters for ludwigite at 320K and 110K from single crystal x-ray diffraction}
	\label{xray-data-tab}
\end{table}
%%%%%%%%%%%%%%%%%%%%%%%%
%====================================
%%%%%%%%%%%%%%%%%%%%%%%%
%====================================
It can be seen that the main consequence of the phase transition on the structure is that some atomic positions split into two positions which are no longer equivalent by symmetry in the low temperature phase. It is also noteworthy that among the Fe cations, only the Fe4 cations undergo such a splitting, which suggests that the transition can be driven by charge ordering in the Fe4-Fe2-Fe4 ladders along which one electron is delocalized at high temperature. This is clearly visible in Figure \ref{ladderhtbt} : above the transition, the Fe2 cations present a strongly anisotropic atomic displacement parameter (a.d.p.) along the rung, while below it, this a.d.p. drastically decreases and the Fe2 position is displaced toward Fe4a or Fe4b in an alternate way along the c axis. In order to characterize this effect, the ionic valences were calculated using the bond valence sum (b.v.s.) method\cite{bvs}. 
%%%%%%%%%%%%%%%%%%%%%%%%
%====================================
%%%%%%%%%%%%%%%%%%%%%%%%
%====================================
In Figure \ref{transit} are represented as function of temperature the evolution of the Fe-Fe distances along the rungs of the Fe4-Fe2-Fe4 ladders, the Fe cation valences and the Fe2 a.d.p.'s. These parameters were found to display the most prominent modifications across the transition. 
%%%%%%%%%%%%%%%%%%%%%%%%
%====================================
%%%%%%%%%%%%%%%%%%%%%%%%
%====================================
At higher temperature, the Fe2-Fe4 distances is $\approx$ 2.79\AA ; well below the transition, Fe2-Fe4a $\approx$ 2.61\AA, an extremely short distance (the Fe-Fe distance in metallic iron is $\approx$ 2.48\AA) while Fe2-Fe4b increases to $\approx$ 2.95\AA. At the same time, the valences of Fe4a and Fe4b become different. Above the transition, the Fe4 cation valence is found to be +2.67, which is the expected value if one extra electron is shared in a disordered way among the three $Fe^{3+}$ cations (2 Fe4 and 1 Fe2) which constitute the ladder rung. However, the Fe2 cation valence is found to be close to +2.5. This low value, together with the large and anisotropic a.d.p. for Fe2 strongly indicate that this cation is in a disordered position above the transition, very probably dynamic, i.e. the cation vibrates between to positions symmetric about the rung center, coming alternately close to the two Fe4 cations at the rung extremities. 

Below the transition, ordering occurs and the Fe2 displacements toward the rung extremities (Fe4a cations) alternate in a zigzag way along the c axis. At the same time the Fe2 a.d.p.'s become isotropic and similar to those of the other atoms. The Fe4a valence decreases to $\approx$+2.53, close to the Fe2 cation valence, while the Fe4b valence increases to $\approx$+2.86. Clearly, this structural phase transition is related to the ordering of charges between Fe4a and Fe4b. A schematic picture can be proposed where the Fe2 and Fe4a cations form a pair by sharing the common electron which was delocalized over the whole rung in the high temperature phase, while the Fe4b valence becomes +3. However, as indicated by the anomalous behavior of the Fe2 cations, the Fe2-Fe4 pairs probably already exist in the high temperature phase but are dynamically disordered between the two sides of the rung. Therefore, the transition can also be viewed as an ordering of the pairs or of the electrons which are responsible for their existence. On cooling towards the transition, the decrease of the available thermal energy could be the driving force to induce the localization of the Fe2 cations (or Fe2-Fe4 pairs). The zigzag order along the c axis of the Fe2-Fe4 pairs would be a way to minimize the structural distortions effects induced by this localization. 

As can be seen in Figure \ref{transit}, the phase transition takes place over a quite wide temperature range of more than 50K before all parameter values are stabilized. This indicates that a large amount of disorder is still present down to $\approx$ 200K. Furthermore, the values at which the evolution of the Fe cation valences saturate do not correspond to complete charge ordering even down to 110K. This will have important consequences for the magnetic behavior of this compound, since the magnetic ordering process will take place on a context of dynamical charge disorder, at least within the Fe4-Fe2-Fe4 ladders.   

\subsection{Neutron powder diffraction experiment\label{NPD}}

In order to investigate the evolution of the magnetic ordering for Fe$_3$BO$_5$, a neutron powder diffraction (NPD) experiment was performed on the D20 instrument of the I.L.L. The 5g sample was put in a cylindrical vanadium can inside an orange cryostat. A wavelength of 1.3\AA  from the Cu(200)monochromator reflection was used, which corresponds to the optimized flux configuration of the instrument. The sample was first heated to 320K, that is above the structural phase transition, and then cooled at an approximately constant rate of 1K/9 minutes down to 10K while diffractograms were continuously measured every 10 minutes. Each diffractogram covers a temperature range of $\approx1.1$K. The data were analyzed using the program Fullprof\cite{fullp}. The structural phase transition is almost undetectable with these data. The only observable though very weak superstructure reflections are the (1 3 1/2) and (2 2 1/2) at about 23.4\degre.

%%%%%%%%%%%%%%%%%%%%%%%%
%====================================
\begin{figure}[btp]
	\begin{center}
		\includegraphics[width=0.4\textwidth]{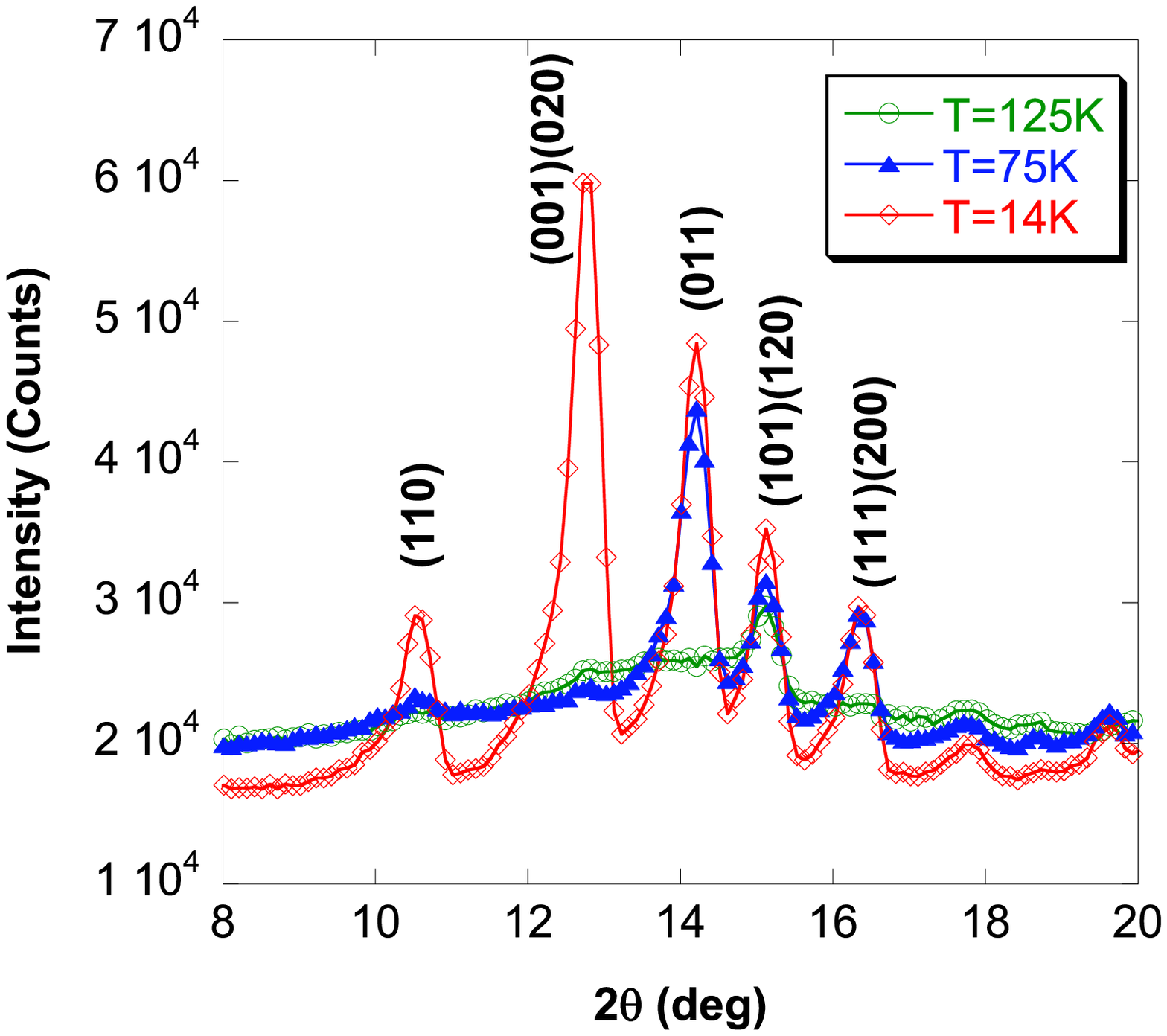}
		\caption{Low angle part of the neutron diffractograms for Fe$_3$BO$_5$ at 125K, 75 and 14K.  The two different sets of magnetic peaks corresponding to the two ordering transitions are clearly visible, as well as magnetic diffuse scattering at the higher temperatures.}
		\label{temp-var1}
	\end{center}
\end{figure}
%%%%%%%%%%%%%%%%%%%%%%%%
%====================================
\begin{figure}[btp]
	\begin{center}
		\includegraphics[width=0.4\textwidth]{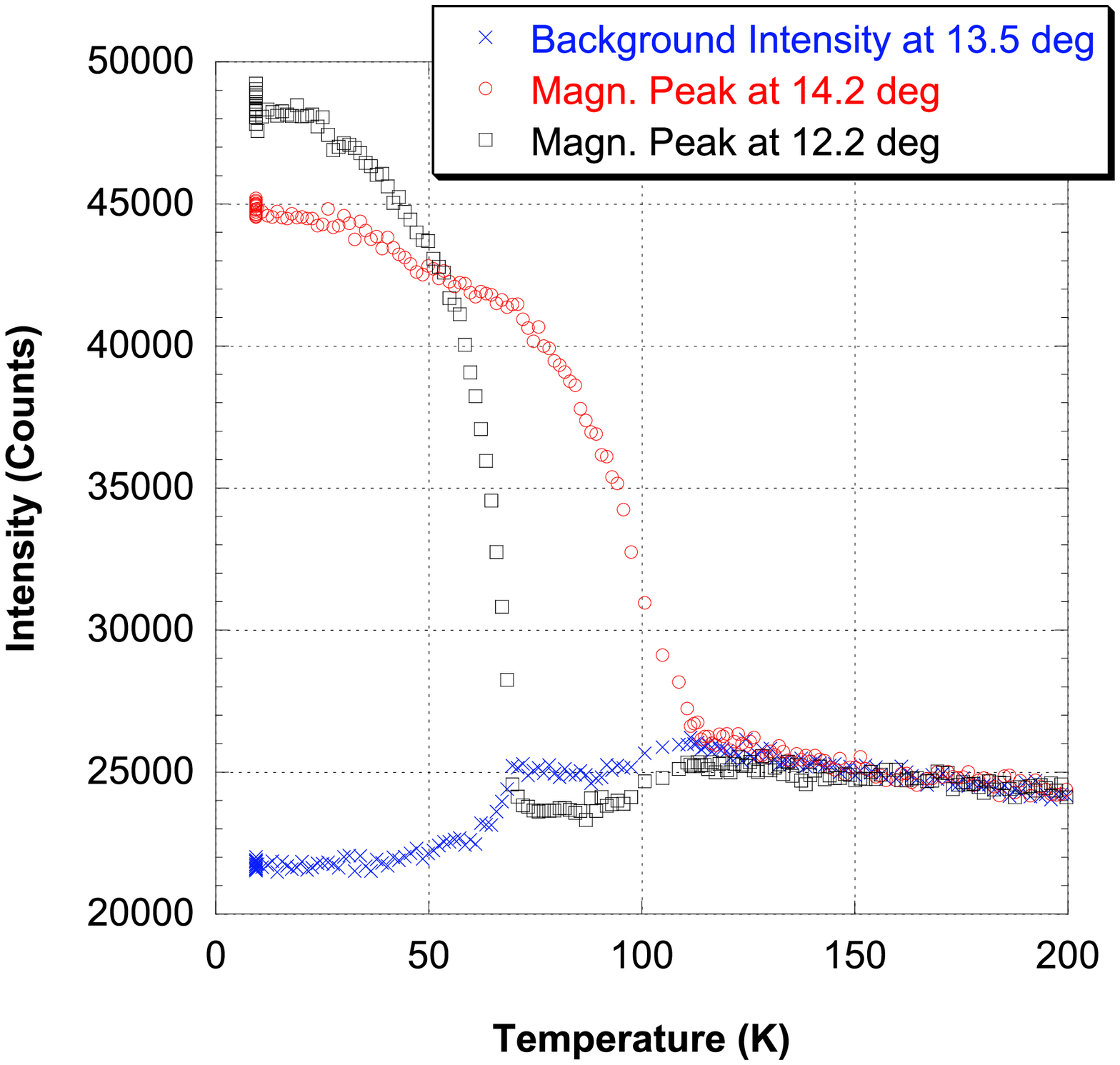}
		\caption{Evolution with temperature of the diffracted intensity for Fe$_3$BO$_5$ at three positions corresponding to background (crosses), higher temperature magnetic phase (circles) lower temperature magnetic phase (squares).}
		\label{temp-var2}
	\end{center}
\end{figure}
%%%%%%%%%%%%%%%%%%%%%%%%
The thermal evolution of the low angle part of the neutron diffractograms is shown in Figure \ref{temp-var1}. The evolution of the diffracted intensities at three selected angles are shown in Figure \ref{temp-var2}. The chosen angles correspond to the positions of two magnetic peaks and of a background point in between them. At high temperatures, a broad hump centered at about 14\degre is clearly visible in the background in Figure \ref{temp-var1}. Its intensity increases steadily with decreasing temperature, as shown in Figure \ref{temp-var2}. At 115 K, a first set of magnetic peaks appears, accompanied by a small decrease of background intensity. At 70K new magnetic peaks set in, and the background decreases to its high angle value. In the whole temperature range, the magnetic peaks can be indexed on the basis of the crystallographic unit cell. These observations indicate that the compound undergoes two successive magnetic ordering transitions on cooling. Since a strong diffuse scattering of magnetic origin persists between the two transitions, the higher temperature magnetic order must concern only part of the spin system, the rest remaining disordered down to the second transition. The two magnetic structures were solved and refined using data at 82K and 10K, and their evolutions were followed by sequential refinement using Fullprof. The treatments included a full structural refinement of the crystal structure, with a single isotropic a.d.p. for each atom type. The peak shapes were described with a pseudo-Voigt function with reflection widths following the Cagliotti law. The peak asymmetry evidenced in figure \ref{temp-var1} is essentially due to axial divergence of the experimental setup. It was taken into account in the Rietveld refinement using the Berar empirical model \cite{berar}. 
%%%%%%%%%%%%%%%%%%%%%%%%
%====================================
\begin{table}[!hb]
	\begin{center}
		\begin{tabular}{|c|c|c|c|}
		\hline
		 \multicolumn{4}{|c|}{Magnetic Structure at T=82K}\\
		\hline
		Atom & Mx & My & Mz \\
		\hline
		\hline
		Fe2 & 0 & 2.3(1) & 0 \\
		\hline
		Fe4a & 0 & 2.3(2) & 0 \\
		\hline
		Fe4b & 0 & 2.4(2) & 0 \\
		\hline
		\multicolumn{4}{|c|}{Magnetic Structure at T=10K}\\
		\hline
		Atom & Mx & My & Mz \\
		\hline
		Fe1 & 3.3(2) & 0 & 0\\
		\hline
		Fe2 & 0 & 3.9(1) & 0\\
		\hline
		Fe3 & 4.0(1) & 0 & 0\\
		\hline
		Fe4a & 0 & 2.74(7) & 0\\
		\hline
		Fe4b & 0 & 2.74 & 0\\
		\hline
		\end{tabular}
	\end{center}
	\caption{Magnetic Structure parameters for ludwigite at 82K (Rp=6.57, Rwp=6.61, Rexp=1.44, Chi2=21.2, RBragg=1.8, Rmag=6.6) and 10K (Rp=7.86, Rwp=8.18, Rexp=1.31, Chi2=39.1, RBragg=2.6, Rmag=5.3).}
	\label{magstru}
\end{table}
%%%%%%%%%%%%%%%%%%%%%%%%
%====================================
\begin{figure}[btp]
	\begin{center}
		\includegraphics[width=0.4\textwidth]{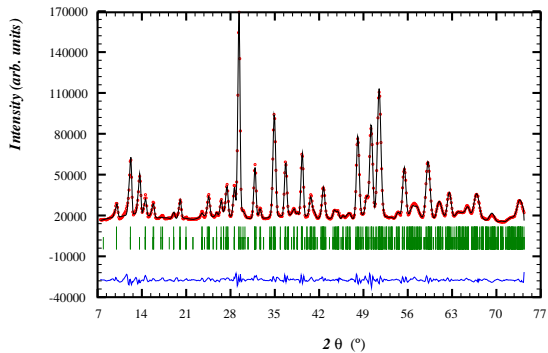}
		\caption{Rietveld plot of the NPD data refinement at 10K for Fe$_3$BO$_5$. The tick marks correspond to the nuclear (top) and magnetic (bottom) reflection positions.}
		\label{riet}
	\end{center}
\end{figure}
%====================================
\begin{figure}[btp]
	\begin{center}
		\includegraphics[width=0.5\textwidth]{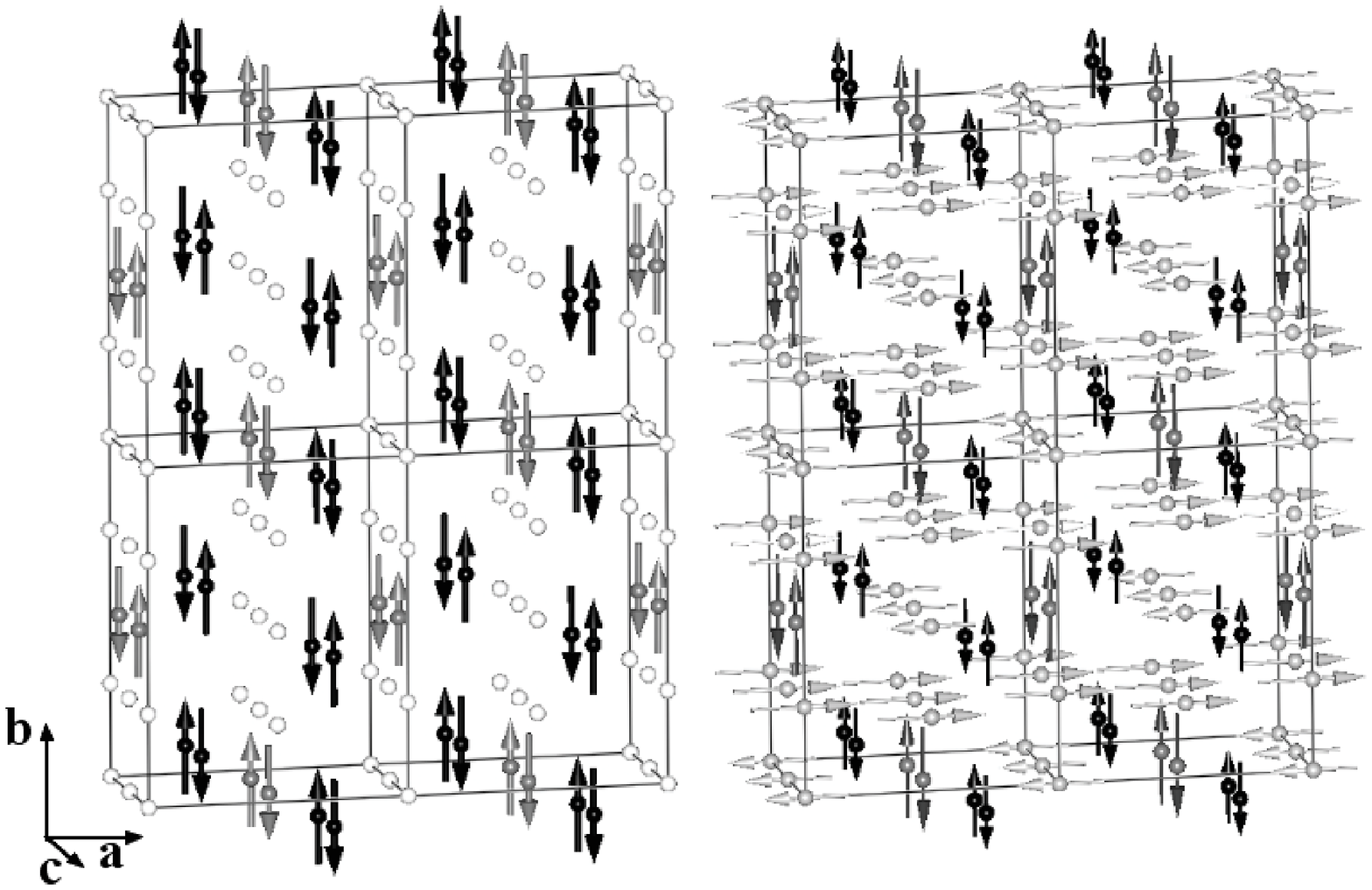}
		\caption{Magnetic structure at 82K (left) and 10K (right) for Fe$_3$BO$_5$. Only the Fe atoms are shown. At 82K : Fe1 and Fe3 : white (no moment); Fe2 : gray, Fe4a, Fe4b : black. At 10K : Fe1 and Fe3 : light gray; Fe2 : dark gray, Fe4a, Fe4b : black}
		\label{HTMS}
	\end{center}
\end{figure}
%====================================
\begin{figure}[btp]
	\begin{center}
		\includegraphics[width=0.4\textwidth]{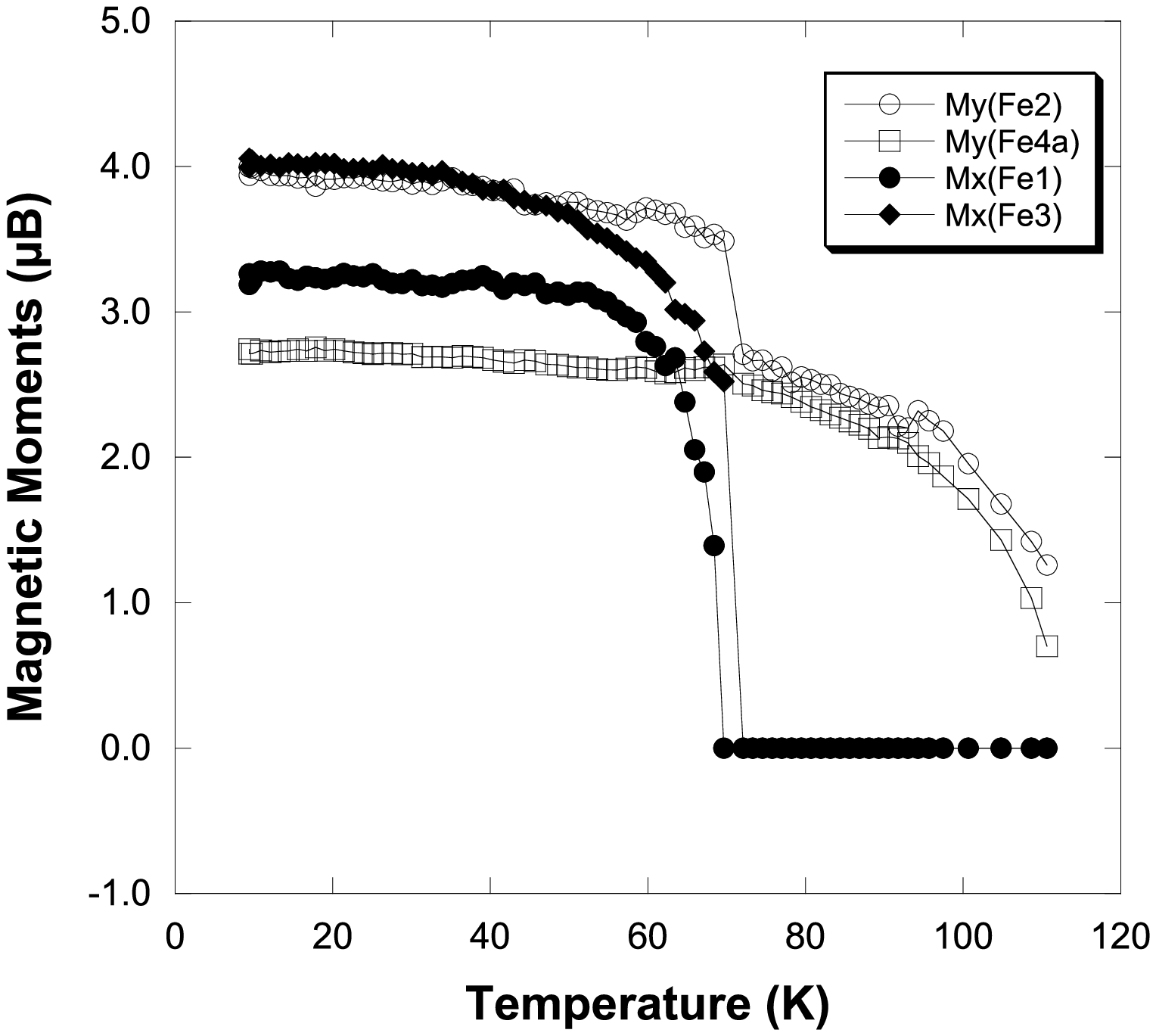}
		\caption{Evolution with temperature of the magnetic moments for Fe$_3$BO$_5$ from Rietveld refinement of NPD data. The moduli of Fe4a and Fe4b are constrained to be equal at all temperatures.}
		\label{mom-vs-t}
	\end{center}
\end{figure}

The results of the magnetic structure refinements at these two temperatures are given in Table \ref{magstru}. The Rietveld plot of the 10K refinement is shown in Figure \ref{riet}. The high temperature magnetic structure (HTMS) between 70K and 115K was refined to Rmag=6.6\%. It involves only the Fe2, Fe4a and Fe4b cations from ladder I. As shown in Figure \ref{HTMS}, it consists in ferromagnetically coupled antiferromagnetic chains running along the c axis within each ladder. All moments are aligned along the b axis. The refined values of the moments on Fe2, Fe4a and Fe4b were equal at 2.3(1)$\mu$B. During sequential refinement as function of temperature, a more stable behavior was obtained by constraining the moments on Fe4a and Fe4b to be equal. The highest moment value obtained just above the second transition were 2.7(1)$\mu$B for Fe2 and 2.51(7)$\mu$B for Fe4a and Fe4b. 
The low temperature magnetic structure (LTMS)is induced by the ordering of ladders II. As shown in Figure \ref{HTMS}, ladders II are formed by antiferromagnetically coupled ferromagnetic chains running along c, all moments being aligned along the a axis. The evolution of the ordered magnetic moments is depicted in Figure \ref{mom-vs-t}. At 10K, the Fe1 and Fe3 moment values are 3.3(2)$\mu$B and 4.0(1)$\mu$B, respectively. The spin arrangement on ladders I remains similar to the HTMS one, but the transition is accompanied by an increase of the Fe2 moments to 3.9(1)$\mu$B at 10K, while the Fe4a and Fe4b moments stay practically constant at 2.7(1)$\mu$B down to the lowest temperatures. This arrangement is close to the one reported by Attfield et al., but in their case all moments were considered as equal. If we try to refine our data with the same constraint, the refinement becomes clearly worse, with Rmag$\approx$12\% compared to 5.3\%.  We can also obtain slightly better refinements by allowing some of the moments to rotate in the (a,b) plane. Only Fe4b and Fe3 rotations out of their direction along the b, resp. a axis yield an improvement of Rmag by $\approx$1\%. However, these positions were quite unstable as function of temperature. We consider that they are not well demonstrated by the refinements and that the most probable structure is the one described in Figure \ref{HTMS}.
In order to check the consistency of the magnetic structures determined experimentally with the space group symmetry, we performed a group theory analysis based on the method of Bertaut \cite{bertaut}, using the Basireps software \cite{basireps}. For the $Pbnm$ space group, there are 8 one-dimensional real irreducible representations for the little group Gk associated to the observed magnetic propagation vector (0,0,0). For sites Fe2, Fe4a and Fe4b, the magnetic arrangement is given by the basis vectors of the representation 4 : (u,v,0), (-u,-v,0), (-u,v,0), (u,-v,0), where only the component along the b axis is found non-zero experimentally. For sites Fe1 and Fe3, the magnetic arrangement is given by the basis vectors of the representation 5 : (u,v,w), (u,-v,w), (u,-v,-w), (u,v,-w),[(u,v,w), (u,-v,w), (u,-v,-w), (u,v,-w)]. This corresponds to a ferromagnetic component along the a axis. It could also imply an antiferromagnetic component in the (b,c) plane which is not observed experimentally. In the case of a magnetic structure with several magnetic sites (here 5 sites corresponding to the 5 non-equivalent Fe cations), the onset of the magnetic transitions depends on the relative strengths of the intra-site and inter-site interactions. In the present case, the Fe2, Fe4a, and Fe4b cations order first with basis vectors belonging to the same representation. This means that the inter-site interactions are dominant between these sites forming ladder I. The situation is similar for the Fe1 and Fe3 cations on ladder II which order at lower temperature with the basis vector of the same representation (but different from the one for ladder I). These different transition temperatures and representations for to the two ladders indicate that the magnetic interactions between them are weaker than those prevailing within each of them.

%%%%%%%%%%%%%%%%%%%%%%%%
%====================================
\section{Discussion \label{discussion}}
The single crystal x-ray diffraction analysis down to 110K indicates in ladder I an electron localization and Fe2-Fe4a pair formation with a very short Fe-Fe bond length. In view of the calculated bond valences, it seems however that the electron remains delocalized between the Fe2 and Fe4a cations. The 15K data reported by Mir et al.\cite{mir2} are consistent with this picture. However the presence of a single data point at 15K prevents from detecting a possible further electronic localization correlated to the magnetic ordering transitions. Our neutron powder diffraction data, aimed at solving the magnetic structures, do not have sufficient resolution to tackle this question and are hindered by the presence of magnetic peaks.
The present NPD experiment shows that a considerable dynamic magnetic scattering is already present a high temperature down to the 70K transition. This indicates the existence of strong magnetic correlations at temperatures  4 times higher than the ordering temperature. The magnetic ordering transition at 112K concerns only the Fe cations in ladders I, while ladders II remain disordered and still display very strong magnetic correlations. This is a direct proof of the very weak magnetic coupling between the two types of ladders in this system. In ladders I, the coupling along a rung is ferromagnetic, with negligible difference of moments between the 3 Fe cations. The rungs are antiferromagnetically coupled along the ladder (and c axis) direction. This is in contrast to the theoretical predictions of Whangbo et al.\cite{whangbo} based on the Strongly Interacting Spin Units approach, who proposed antiferromagnetic coupling in the ladders rungs. Recently, Vallejo et al.\cite{vallejo} have proposed an interpretation taking into account the prominent role of itinerant electrons in this system, which was able to predict the right magnetic structure. This is also supported by the relatively moderate values of the resistivity ($\approx$10 $\Omega$.cm at room temperature\cite{mir1}) which is only slightly modified by the charge ordering transition. The value of 2.3$\mu$B for the Fe cation moments is much smaller than what is expected for Fe cations (5 and 4 $\mu$B for $Fe^{3+}$ and $Fe^{2+}$, respectively). This may be another indication that magnetic and electronic ordering is not complete even in ladders I in the 70K-112K temperature range. 
Below 70K, a new magnetic order sets in, where the Fe cations of ladders II form ferromagnetic chains along the c axis which are coupled antiferromagnetically. The magnetic order on ladders I is not modified, except for an increase of the Fe2 moments. Examining Figure \ref{HTMS}, one can notice that the Fe2 antiferromagnetic chains of ladders I are weakly connected to ladders II through two Fe3 ferromagnetic chains. Therefore, this Fe2 moment increase is probably not a consequence of the magnetic ordering of ladders II. It might be related to a further ordering of charges between the Fe2 and Fe4a cations, but as stated above, we do not have any direct proof of this effect yet. The difference of magnetic coupling along the rung of ladders I and ladders II probably originates in the strong structural difference between the two types of ladders. The Fe2, Fe4a and Fe4b cations of ladders I are connected along a rung via edge-sharing at distances between 2.6 and 2.95\AA. The Fe1 and Fe3 cations of ladders II are connected via corner sharing at distances above 3.35\AA. At 10K, the Fe1 and Fe3 moment values (3.3 and 3.9$\mu$B) are reasonable for a $Fe^{2+}$ cation, but those for Fe2, Fe4a and Fe4b cations of ladders I are still small compared to expected values. Since there are twice as many Fe3 than Fe1 cations per cell, the magnetic arrangement leads to a net moment along the a axis of 18.9$\mu$B per cell (i.e. 0.79$\mu$B per Fe cation). Indeed, the presence of ferromagnetism was demonstrated in this compound below 70K\cite{guimar} by magnetization measurements. However, the same authors observed its disappearance below 40K. We do not detect any anomalous change in the NPD diffractograms around this temperature. If the magnetic structure is preserved, the disappearance of the net magnetic moment along the a axis would be possible only if the magnitude of the Fe1 moments becomes twice that of the Fe3 moments, which would certainly be detectable with our data. The origin of this magnetization change at 40K as to be searched elsewhere.

%%%%%%%%%%%%%%%%%%%%%%%%
%====================================
\section{Summary \label{summary}}
In this study, we have used single crystal x-ray diffraction and neutron powder diffraction as function of temperature to investigate the charge and magnetic ordering in Fe$_3$BO$_5$ ludwigite. A charge ordering transition occurs close to room temperature. X-ray diffraction indicates that it takes place over a wide temperature range between 200K and 300K. It consists in an order-disorder transition within one of the two three-leg ladders contained in the structure and is related to the localization of an electron to form a pair of iron cations within the ladder's rung. Above the transition this electron is dynamically delocalized over the three iron cations forming the rung. To minimize the effects of the structural distortion created by this electron localization, the iron cation pairs order in a zig-zag way along the ladder's direction. At lower temperatures, ludwigite has been reported to undergo two successive antiferromagnetic and ferromagnetic ordering transitions at 112K and 70K, respectively. Refinements of neutron powder diffraction data indicate that the first transition is due to magnetic ordering of the ladders which charge-order around room temperature, forming ferromagnetically coupled antiferromagnetic chains running along the ladder axis. At 70K, magnetic order of the second type of ladders sets in. It consists in antiferromagnetically coupled ferromagnetic chains running along the ladder axis. This ferrimagnetic arrangement remains unchanged down to 10K and leads to a net magnetic moment of  $\approx$19$\mu$B per cell at low temperature. We observe no modification of the NPD diffractograms around 40K, where the ferromagnetic component has been reported to disappear.

\begin{acknowledgments}
The authors would like to thank V. Simonet, J. Dumas, M. Avignon and M. Continentino for fruitful discussions, J. Fernandes for providing us with the samples and T. Hansen for his help during the D20 experiment. 
\end{acknowledgments}

%\%%%%%%%%%%%%%%%
%====================================


\begin{thebibliography}{1}

\bibitem{conti}M. A. Continentino, J.C. Fernandes, R.B. Guimaraes, B. Boechat, and A. Saguia,  in  "Frontiers in Magnetic Materials", Vol. 24, p.385, ed. A.V. Narlikar, Springer Publ. (2005),  

\bibitem{guimar}R.B. Guimaraes, M. Mir, J.C. Fernandes, M.A. Continentino, H.A. Borges, G. Cernicchiaro, M.B. Fontes, D.R.S. Candela,  E. Baggio-Saitovitch, Phys. Rev. {\bf B60}, 6617 (1999).

\bibitem{larrea}J. Larrea, D.R. Sanchez, F.J. Litterst, E. Baggio-Saitovitch, J.C. Fernandes,  R.B. Guimaraes, M.A. Continentino, Phys. Rev. {\bf B70}, 174452 (2004);  J. Larrea, D.R. Sanchez, F.J. Litterst, E. Baggio-Saitovitch, J. Phys. Cond. Mat. {\bf 13}, L949 (2001);  J. Larrea, D.R. Sanchez, E. Baggio-Saitovitch, J.C. Fernandes,  R.B. Guimaraes, M.A. Continentino,  F.J. Litterst, J. Magn. Mag. Mat. {\bf 226-230}, 1079 (2001). 

\bibitem{mir1}M. Mir, R.B. Guimaraes, J.C. Fernandes et al.,  Phys. Rev. Lett. {\bf 87}, 147201 (2001). 
\bibitem{mir2}M. Mir, J. Janczak, Y.P. Mascarenhas, J. of Appl. Cryst. {\bf 39}, 42 (2006)
\bibitem{latge}A. Latge, M.A. Continentino, Phys. Rev. {\bf B66}, 094113 (2002).
\bibitem{whangbo}M.H. Whangbo, H.J. Koo, J. Dumas, M.A. Continentino, Inorg. Chem. {\bf 41}, 2193 (2002).
\bibitem{vallejo}E. Vallejo, M. Avignon, Phys. Rev. Lett. {\bf 97}, 217203 (2006); E. Vallejo, M. Avignon, J. Magn. Magn. Mat. {\bf 310}, 1130 (2007).
\bibitem{attfield}J. Paul Attfield, John F. Clarke and David A. Perkins, Physica {\bf B 180 \& 181} (1992) 581-584
\bibitem{bordet1}P. Bordet , E. Suard, Acta Cryst. {\bf A61}, C57 (2005).
\bibitem{bordet2}P. Bordet, B. Anterion, M. Mir, R.B. Guimaraes, J.C. Fernandes, M.A. Continentino, Acta Cryst. {\bf A58 (Suppl)}, C363, (2002).
\bibitem{denzo}Z. Otwinowski and W. Minor In Methods in Enzymology, 276,edited by C.W. Carter, Jr. \& R.M. Sweet pp. 307-326, (1997)New York:Academic Press..
\bibitem{berar}J.-F. Bérar \& G. Baldinozzi, J. Appl. Cryst. {\bf 26}, 128 (1993).
\bibitem{maxus}S. Mackay, C.J. Gilmore, C. Edwards, N. Stewart \& K. Shankland,(1999). maXus  Computer Program for the Solution and Refinement of Crystal Structures. Bruker Nonius, The Netherlands, MacScience, Japan \& The University of Glasgow.
\bibitem{jana}V. Petricek, M. Dusek \& L. Palatinus,(2000). Jana2000. The crystallographic computing system. Institute of Physics, Praha, Czech Republic. 
\bibitem{mir}M. Mir, R.B. Guimaraes, J. C. Fernandes, M. A. Continentino, A. C. Doriguetto, Y. P. Mascarenhas, J. Ellena, and E. E. Castellano, R. S. Freitas and L. Ghivelder, Phys. Rev. Lett. {\bf 87}, 147201 (2001).
\bibitem{bvs} N.E. Brese \& M. O'Keeffe, Acta Cryst. {\bf B47}, 192 (1991).
\bibitem{fullp} J. Rodriguez-Carvajal, Physica B{\bf 192}, 55 (1993).
\bibitem{bertaut} E. F. Bertaut, Acta Cryst. A24, {\bf A24}, 217 (1968).
\bibitem{basireps} J. Rodriguez-Carvajal, program available with the fullprof distribution
\end{thebibliography}
\end{document}